\documentclass[a4paper,10pt]{article}
\usepackage{graphicx}
\usepackage{comment}
\usepackage{amssymb}
\usepackage{amsmath}
\usepackage{nicefrac}
\usepackage{graphicx}
\usepackage{dcolumn}
\usepackage{bm}
\usepackage{comment}
\usepackage{multirow}
\usepackage{cite}
\usepackage{mathrsfs}

\begin{document}

\newcommand{\bin}[2]{\left(\begin{array}{c}\!#1\!\\\!#2\!\end{array}\right)}
\newcommand{\threej}[6]{\left(\begin{array}{ccc}#1 & #2 & #3 \\ #4 & #5 & #6 \end{array}\right)}
\newcommand{\sixj}[6]{\left\{\begin{array}{ccc}#1 & #2 & #3 \\ #4 & #5 & #6 \end{array}\right\}}
\newcommand{\regge}[9]{\left[\begin{array}{ccc}#1 & #2 & #3 \\ #4 & #5 & #6 \\ #7 & #8 & #9 \end{array}\right]}
\newcommand{\La}[6]{\left[\begin{array}{ccc}#1 & #2 & #3 \\ #4 & #5 & #6 \end{array}\right]}
\newcommand{\hj}{\hat{J}}
\newcommand{\hux}{\hat{J}_{1x}}
\newcommand{\hdx}{\hat{J}_{2x}}
\newcommand{\huy}{\hat{J}_{1y}}
\newcommand{\hdy}{\hat{J}_{2y}}
\newcommand{\huz}{\hat{J}_{1z}}
\newcommand{\hdz}{\hat{J}_{2z}}
\newcommand{\hup}{\hat{J}_1^+}
\newcommand{\hum}{\hat{J}_1^-}
\newcommand{\hdp}{\hat{J}_2^+}
\newcommand{\hdm}{\hat{J}_2^-}
\newcommand{\va}{{\bf A}}
\newcommand{\vl}{{\bf L}}
\newcommand{\vm}{{\bf M}}
\newcommand{\vn}{{\bf N}}
\newcommand{\vr}{{\bf r}}
\newcommand{\vj}{{\bf j}}
\newcommand{\vF}{{\bf F}}
\newcommand{\vp}{{\bf p}}
\newcommand{\valpha}{{\bf \alpha}}
\newcommand{\vbeta}{{\bf \beta}}

\huge

\begin{center}
Sum rules for Clebsch-Gordan coefficients from group theory and Runge-Lenz-Pauli vector
\end{center}

\vspace{0.5cm}

\large

\begin{center}
Jean-Christophe Pain$^{a,b,}$\footnote{jean-christophe.pain@cea.fr}
\end{center}

\normalsize

\begin{center}
\it $^a$CEA, DAM, DIF, F-91297 Arpajon, France\\
\it $^b$Universit\'e Paris-Saclay, CEA, Laboratoire Mati\`ere en Conditions Extr\^emes,\\
\it 91680 Bruy\`eres-le-Ch\^atel, France
\end{center}


\begin{abstract}
We present sum rules for Clebsch-Gordan coefficients in the framework of $SO(4)$ group-theoretical description of the hydrogen atom. The main results are obtained using properties of the Runge-Lenz-Pauli vector, in particular expressing the matrix elements of the powers of its last component both in spherical and parabolic basis. Connections with Stark effect and diamagnetism of the hydrogen atom are outlined. 
\end{abstract}

\section{Introduction}\label{sec1}

A number of sum rules involving Wigner $3jm$ or Clebsch-Gordan coefficients have been discovered in the framework of atomic, molecular and nuclear spectroscopy \cite{Herrick1975,Berghe1976,Rashid1976,Barut1976,Morgan1976,Morgan1977,Klarsfeld1978,Meyer1978,Kulesza1980,Labarthe1980,Din1981,Askey1982,Ancarani1993,Ancarani1994,Stehle1996,Casini1997,Mosse2004,Gigosos2007,Stambulchik2008,Gilleron2019,Pain2019}. They are related to different topics such as Stark and Zeeman effects, electron impact theory, charge transfer cross-sections, molecular collisions, etc. Sum rules reveal information about a physical problem, since they are associated to invariance properties and therefore reflect the symmetries of the system. They can also be useful for checking numerical computations involving angular-momentum coupling. When they are connected to a basis change, they are likely to help finding the best mathematical description to make the calculations as simple as possible.

Besides the angular momentum $\vl=\vr\times \vp$, there is another conservative operator which is the Runge-Lenz vector $\va_{RL}={\bf L}\times{\bf p}+{\bf r}/r$ \cite{Runge1919,Lenz1924} (throughout the paper we use atomic units $m=\hbar=e=1$ and omit the hat conventionally denoting quantum-mechanical operators). In its classical interpretation, $\va_{RL}$ points in the direction of the semi-major axis of the elliptical Kepler orbit and its magnitude is equal to the ellipse's eccentricity. $\va$ was chosen according to the first letter of the German word ``Achsenvektor'' \cite{Lenz1924} as a signature of its ``axial vector'' character \cite{Hey2015}, meaning that the vector is directed along the major axis of the conic section orbit. In Ref. \cite{Hey2019}, Hey pointed out that this translation may not be appropriate, since the term ``axial vector'' is now conventionally used as a synonym for a ``pseudovector'', in contrast to a ``polar vector'', well suited for $\va$. Hey suggested that a better translation would be ``apsidal'' vector.

The existence of closed orbits is a consequence of the conservation of $\va_{RL}$, \emph{i.e.} $d\va_{RL}/dt=0$ \cite{Rau1990}. The classical expression of the Runge-Lenz vector needs to be symmetrized in order to ensure the hermiticity of the corresponding quantum mechanical operator $\va_{RLP}$:
\begin{equation}
\va_{RLP}=\frac{1}{2}\left({\bf L}\times{\bf p}-{\bf p}\times{\bf L}\right)+\frac{\bf r}{r}. 
\end{equation}
The outline of a derivation provided by Pauli \cite{Pauli1926} (for that reason we add the letter $P$ to the subscript) was supplemented by detailed calculations by Becker and Bleuler \cite{Becker1976}, Flamand \cite{Flamand1966} and Valent \cite{Valent2003}, involving commutator algebra as well as angular-momentum Lie algebra. As mentioned by many authors (see for instance Refs. \cite{Hey2015} and \cite{Chua2018}), the Runge-Lenz-Pauli vector is a very versatile tool, which can provide recurrence relations for expectation values of powers of operators and an alternative proof of the Kramers-Pasternack relation for instance. This operator is of great interest in Stark- and Zeeman-effect calculations for magnetic-confinement-fusion edge plasmas \cite{Rosato2014} and tenuous space plasmas (H II regions). It brings a number of valuable results for plasma diagnostics requiring rate calculations of atomic processes. For convenience, we define, as commonly made, the reduced Runge-Lenz-Pauli vector $\va=\va_{RLP}/\sqrt{-2E}$, where $E$ is an eigenvalue of 
\begin{equation}
H_0=\frac{p^2}{2}-\frac{1}{r}
\end{equation}
the Hamiltonian of the hydrogen atom. 

The supersymmetry of the Coulomb problem has been recognized for a long time \cite{Kustaanheimo1965,Bander1966a,Bander1966b,Decoster1970a,Decoster1970b,Boiteux1972,Englefield1972,Boiteux1973a,Boiteux1973b,Barut1973,Chen1985,Maclay2020}. Historically, in his pioneering work, Pauli did not explicitly identify $\vl$ and $\va$ as generators of the symmetry group $SO(4)$. In 1935, Fock \cite{Fock1935} showed that the momentum ($p$) space wave functions of bound states were spherical harmonics in four dimensions. Fock stated that four-dimensional rotations reflect the symmetry of the degenerate bound energy levels in momentum space, thus realizing $SO(4)$, which is the group of special orthogonal transformations leaving the norm of a four-dimensional vector constant. By numbering the four-dimensional spherical harmonics $Y_{n\ell m}$ in momentum space ($-\ell\leq m\leq\ell$, where $0\leq\ell\leq n-1$), he found that the degeneracy of an energy level identified by the principal quantum number $n$ is $n^2$. A few months later, Bargmann showed that the angular momentum $\vl$ and the Runge-Lenz-Pauli vector $\va$, obey the commutation rules of $SO(4)$ \cite{Bargmann1936}. Bargmann pointed out the close connection between the preservation of Runge-Lenz-Pauli vector and solutions of Schr\"odinger's equation in parabolic coordinates.

Both $\vl$ and $\va$ commute with $H_0$. The commutation relations of $L_i$ and $A_i$ ($i=1,2,3$ or $x,y,z$), forming the algebra of $O(4)$ \cite{Bander1966a}, are identical to the ones obeyed by the classical Poisson brackets:
\begin{equation}
\left[L_i,L_j\right]=i\epsilon_{ijk}L_k,\;\;\;\;\left[L_i,A_j\right]=i\epsilon_{ijk}A_k\;\;\;\;\left[A_i,A_j\right]=i\epsilon_{ijk}L_k,
\end{equation}
where $\epsilon_{ijk}$ is the usual Levi-Civita tensor:
\begin{equation}
\epsilon_{ijk}=\left|
\begin{array}{ccc}
\delta_{i,1} & \delta_{i,2} & \delta_{i,3}\\
\delta_{j,1} & \delta_{j,2} & \delta_{j,3}\\
\delta_{k,1} & \delta_{k,2} & \delta_{k,3}
\end{array}\right|,
\end{equation}
which can be formulated as
\begin{equation}
\epsilon_{ijk} = \left\{
\begin{array}{cc}
+1 & \mbox{if } (i,j,k) \mbox{ is } (1,2,3), (2,3,1) \mbox{ or } (3,1,2), \\
-1 & \mbox{if } (i,j,k) \mbox{ is } (3,2,1), (1,3,2) \mbox{ or } (2,1,3), \\
0 & \mbox{if }i=j \mbox{ or } j=k \mbox{ or } k=i.
\end{array}
\right.
\end{equation}
The Coulomb problem is equivalent to a pair of two-dimensional harmonic oscillators subject to a constraint, which constitutes the so-called oscillator representation \cite{Kustaanheimo1965,Bander1966a,Bander1966b,Boiteux1972,Boiteux1973a,Boiteux1973b,Barut1973,Chen1985,Maclay2020}. In their two seminal papers about the $O(4)$ symmetry of the hydrogen atom, Bander and Itzykson provide an introduction to the spectrum generating group (also called ``non-invariance group'') $SO(4,1)$ \cite{Bander1966a,Bander1966b}. Some generators of the Lie algebra of $SO(4,1)$ move the states from one energy level to another. Bander and Itzykson establish the equivalence of the approach based on $\vl$ and $\va$ forming the $O(4)$ group, and the one relying on the solutions in parabolic coordinates. They point out that the stereographic projection, first introduced by Fock and drawing the correspondence between the momentum space and the position (sometimes called ``configuration'') space depends on the energy. This implies that the $SO(4)$ subgroup is relevant in a subspace of constant energy only. Bander and Itzykson also worked on extensions of the $SO(4)$ group to include the possibility for changing the principal quantum number to another. This requires to build subspaces of $SO(4)$. Each value of $n$ corresponds to a separate $SO(4)$ subgroup. An infinite number of states is required in order to obtain a representation of $SO(4,1)$, which is precisely the case of the hydrogen atom. $SO(4,1)$ was extended to form the non-compact group $SO(4,2)$ (dynamical group of the hydrogen atom) which involves both the symmetry operators of the Hamiltonian and the transition operators \cite{Barut1973,Adams1994}.

In section \ref{sec2}, the group-theoretical formalism is introduced. The matrix elements of the square modulus $A^2$ and the square of the $z$-component of the Runge-Lenz-Pauli vector are provided in section \ref{sec3} both in spherical and parabolic basis. A sum rule involving $\ell(\ell+1)$ weighting factor, already presented in a previous paper \cite{Pain2021}, is obtained in a straightforward way using matrix elements of $A^2$ in section \ref{sec4} and a new sum rule, deduced from matrix elements of $A_z^2$, is presented in section \ref{sec5}. The connection with the Stark effect and the diamagnetism of the hydrogen atom in the low-field limit are discussed in sections \ref{sec6} and \ref{sec7} respectively. A general procedure to derive new identities is explained in the latter section, and two additional sum rules, obtained from matrix elements of $A_z^3$ and $A_z^4$ respectively, are explicitly given. Equations (\ref{sr2}), (\ref{sr3}) and (\ref{sr4}) constitute the main results of the present work.

\section{Runge-Lenz-Pauli vector and group theory}\label{sec2}

In a classical picture, $\va$ and $\vl$ completely define a closed Kepler ellipse. The equivalence between the Coulomb problem and harmonic oscillator is strongly related to the Bertrand theorem, which states that among central-force (radial) potentials with bound orbits, there are only two types of potentials satisfying the property that all bound orbits are also closed. The first one is an inverse square radial force (such as the gravitational or electrostatic potential) and the second one is the quadratic potential of the radial harmonic oscillator \cite{Bertrand1873}. Using the above defined scaled version of the $\va$ vector \emph{i.e.} $\va=(-2E)^{-1/2}\va_{RLP}$, the basic quantum relations are then 
\begin{equation}\label{en}
\va.\vl=0 \;\;\;\;\mathrm{and}\;\;\;\; A^2+L^2+1=-1/(2E).
\end{equation}
The energy levels can be deduced from the determination of the representations of the group $SO(4)$ which are realized by the degenerate eigenstates \cite{Pauli1926,Bargmann1936}. The representations of $SO(4)$ are characterized by the Casimir operators:
\begin{equation}
C_1=\vl.\va\;\;\;\;\mathrm{and}\;\;\;\;C_2=L^2+A^2.
\end{equation}
Once $C_2$ is known, the eigenvalues of $H_0$ follow from Eq. (\ref{en}). In order to determine the possible values of $C_2$, the $O(4)$ algebra can be factored into two disjoint $SU(2)$ algebras, obeying the same commutation relations as the angular momentum operators \cite{Bander1966a}:
\begin{equation}
\vn = (\vl+\va)/2 \;\;\;\;\mathrm{and}\;\;\;\; \vm = (\vl-\va)/2.
\end{equation}
Such commutation relations are actually
\begin{equation}
\left[M_i,N_j\right]=0, \;\;\;\; \left[M_i,M_j\right]=i\epsilon_{ijk}M_k\;\;\;\;\mathrm{and}\;\;\;\;\left[N_i,N_j\right]=i\epsilon_{ijk}N_k.
\end{equation}
The Casimir operators are therefore
\begin{equation}
M^2=j_1(j_1+1)\;\;\;\;\mathrm{and}\;\;\;\; N^2=j_2(j_2+1),
\end{equation}
and the numbers $j_1$ and $j_2$, which may have half-integer values for $SU(2)$ but not $O(3)$, define the $(\vj_1,\vj_2)$ representation of $SO(4)$. From the definition of $\va$ and $\vl$ in terms of the canonical variables, it follow that $C_1=\vl.\va=0$, which means $j_1=j_2=j$, as in the classical case. We obtain
\begin{equation}
M^2=N^2=\frac{L^2+A^2}{4}=j(j+1)
\end{equation}
and therefore
\begin{equation}
L^2+A^2+1=(2j+1)^2,
\end{equation}
$j$ being half-integer. Substituting this result into Eq. (\ref{en}) yields the well-known expression the bound-state energy of the hydrogen atom 
\begin{equation}
E_n=-\frac{1}{2n^2}
\end{equation}
and we have, in the set of the states with fixed $n$ (a so-called $n-$manifold): $\va=n\times\va_{RLP}$. The above considerations about the Casimir operators have shown that the hydrogen atom yields completely symmetrical tensor representations of $SO(4)$, namely $(j,j)=\left(\frac{n-1}{2},\frac{n-1}{2}\right)$, $n=(2j+1)=0,1,2,\cdots$. The dimensionality is $n^2$, corresponding to the $n^2$ (degenerate) states. The operators $\vj_1$ and $\vj_2$ satisfy the relations
\begin{equation}
j_1^2=j_2^2=\frac{n^2-1}{4}.
\end{equation}
This formalism leads to a fully symmetric description of the shell degeneracy in terms of the two independent angular momenta $\vj_1$ and $\vj_2$. One has
\begin{equation}\label{lz}
L_z=j_{1z}+j_{2z}=m.
\end{equation}
The most general symmetry transformation is an exponential of infinitesimal generators
\begin{equation}
\mathcal{T}(\valpha,\vbeta)=\exp\left[-i\left(\valpha.\vl+\vbeta.\va\right)\right]=\exp\left[-i\left(\valpha-\vbeta\right).~\vj_1\right]~\exp\left[-i\left(\valpha+\vbeta\right).~\vj_2\right]. 
\end{equation}
In this form, it appears that a transformation of the symmetry group $SO(4)$ can be written as the product of a rotation generated by $\vj_1$ (belonging to a $SO(3)$ group) by a rotation generated by $\vj_2$ (belonging to a $SO(3)$ group as well). Therefore, a representation of the symmetry group of the hydrogen atom $SO(4)=SO(3)\otimes SO(3)$ can be explicitly given in terms of $\vj_1$ and $\vj_2$. In other words, $\vj_1$ and $\vj_2$ are the generators of the $SO(4)$ algebra for the Coulomb problem. They behave like two independent angular momenta commuting with the Hamiltonian, which allows alternative choices for the operator which can be simultaneously diagonalized with the Hamiltonian to provide the three-dimensional quantization and achieve the complete labeling of states. The particular choice of $L^2$ and $L_z$, that is, of $\left(j_1+j_2\right)^2$ and $j_{1z}+j_{2z}$, leads to the $|n\ell m\rangle$ spherical basis. The group chain for this description is $O(4)\subset O(3)\subset O(2)$ \cite{Rau1990}. An alternative choice of $A_z$ and $L_z$, \emph{i.e.} of $j_{1z}$ and $j_{2z}$ again, gives the parabolic basis with its group chain $O(4)\subset O(3)\subset O(2)\otimes O(2)$.

For more general systems (\emph{i.e.} not hydrogenlike atoms), it is \emph{a priori} not possible to express the level energy in terms of the quantum numbers alone. This can be done in the present case, because the Hamiltonian can be formulated as a function of the Casimir operators depending explicitly on all the quantum numbers. The choice of basis states is not unique. For instance, one can take a basis made with eigenstates of $C_2$, $A_z$ and $L_z$, which is consistent with the use of parabolic coordinates \cite{Bureyeva2001}. Alternatively, one can resort to the spherical basis states $|n\ell m\rangle$ that are eigenstates of $C_2$, $L^2$ and $L_z$. 

\section{Matrix-elements of $A_z$}\label{sec3}

\subsection{Spherical basis}\label{subsec31}

Considering the basis states $|n\ell m\rangle$, one has
\begin{equation}
\sqrt{\left(L^2+A^2+1\right)}|n\ell m\rangle=n|n\ell m\rangle,
\end{equation}
\begin{equation}
L^2|n\ell m\rangle=\ell(\ell+1)|n\ell m\rangle
\end{equation}
as well as
\begin{equation}
L_z|n\ell m\rangle=m|n\ell m\rangle.
\end{equation}
Let us introduce the raising and lowering operators
\begin{equation}
L_{\pm}=L_x\pm iL_y,
\end{equation}
satisfying the commutation relations
\begin{equation}
\left[L^2,L_{\pm}\right]=0\;\;\;\;\mathrm{and}\;\;\;\;\left[L_z,L_{\pm}\right]=\pm L_{\pm}.
\end{equation}
Subsequently, $L_{\pm}$ changes the $m$ value of the basis states
\begin{equation}
L_{\pm}|n\ell m\rangle=\sqrt{\ell(\ell+1)-m(m\pm 1)}~|n\ell (m\pm 1)\rangle
\end{equation}
for $\ell\geq 1$. It is worth mentioning that any $SO(4)$ transformation can be expressed as three successive rotations \cite{Biedenharn1961}: the first induced by $\vl$, the second by $A_z$ and the third by $\vl$. Since $A_z$ commutes with $L_z$ and $C_2$, it modifies only$\ell$:
\begin{eqnarray}\label{nlm}
A_z|n\ell m\rangle&=&\left\{\frac{\left[n^2-(\ell+1)^2\right]\left[(\ell+1)^2-m^2\right]}{4(\ell+1)^2-1}\right\}^{1/2}|n(\ell+1)m\rangle\nonumber\\
& &+\left[\frac{(n^2-\ell^2)(\ell^2-m^2)}{4\ell^2-1}\right]^{1/2}|n(\ell-1)m\rangle.
\end{eqnarray}
Such a relation was also obtained in Ref. \cite{Hey2015} containing an outstanding analysis of the Runge-Lenz-Pauli vector.

\subsection{Parabolic basis states}\label{subsec32}

If we consider the eigenstates of $(H_0,j_{1z},j_{2z})$ with fixed values of $n$ and $m$, there is an isomorphism between this set and the set of eigenstates of the oscillator representation. The two representations are linked through the isomorphism:
\begin{equation}
\begin{array}{l}
j_{1z}+j_{2z}=m\\
j_{1z}-j_{2z}=n_1-n_2\\
j_1^2=j_2^2=j(j+1)\;\;\;\;\mathrm{with}\;\;\;\;n=2j+1,
\end{array}
\end{equation}
where parabolic quantum numbers $n_1$ and $n_2$ satisfy
\begin{equation}
n_1+n_2+|m|+1=n.
\end{equation}
The ladder operators \cite{Infeld1951,Rau1990} act as
\begin{eqnarray}
j_{1+}|n_1n_2m\rangle=\frac{1}{2}\left\{[n-1-m-n_1+n_2)][n+1+m+n_1-n_2]\right\}^{1/2}|(n_1+1)n_2m\rangle& &\nonumber\\
& &
\end{eqnarray}
as well as
\begin{eqnarray}
j_{1-}|n_1n_2m\rangle=\frac{1}{2}\left\{[n-1+m+n_1-n_2)][n+1-m-n_1+n_2]\right\}^{1/2}|(n_1-1)n_2m\rangle,& &\nonumber\\
& &
\end{eqnarray}
and one has
\begin{equation}\label{para}
A_z|n_1n_2m\rangle=(n_1-n_2)|n_1n_2m\rangle.
\end{equation}
The same result was obtained by Hey using the explicit expression of $A_z$ as a function of parabolic variables $\xi$ and $\eta$, as well as derivatives with respect to them \cite{Hey2015}.

\section{Another derivation for the sum rule for rotational moments}\label{sec4}

In a recent work, we presented new sum rules for $3jm$ coefficients, which involve, in addition to the usual weighting factor $(2J + 1)$ where $J$ is an arbitrary angular momentum, the quantity $[J(J+1)]^k$ with $k\geq 1$. Such identities are relevant for instance in the statistical modeling of rotational spectra within the theory of moments, and help determining the expectation values of $r^k$ (used in the theory of Stark effect for hydrogenic ions) in parabolic coordinates from the expectation values of $r^k$ in spherical coordinates \cite{Hey2007,Blaive2009,Pain2021,Hey2021}. One has \cite{Su2017}:
\begin{equation}
\langle n\ell m|A^2|n\ell m\rangle=\langle n\ell m|\left(n^2-1-L^2\right)|n\ell m\rangle=n^2-1-\ell(\ell+1).
\end{equation}
In parabolic coordinates, we have
\begin{equation}
L^2=j_1^2+j_2^2+2\vj_1.\vj_2
\end{equation}
and
\begin{equation}\label{exp}
\vj_1.\vj_2=j_{1x}j_{2x}+j_{1y}j_{2y}+j_{1z}j_{2z}.
\end{equation}
Since
\begin{equation}
j_{1z}|n_1n_2m\rangle=\frac{1}{2}\left(m+n_1-n_2\right)|n_1n_2m\rangle
\end{equation}
and
\begin{equation}
j_{2z}|n_1n_2m\rangle=\frac{1}{2}\left(m-n_1+n_2\right)|n_1n_2m\rangle,
\end{equation}
we get
\begin{equation}
j_{1z}j_{2z}|n_1n_2m\rangle=\frac{1}{4}\left[m^2-\left(n_1-n_2\right)^2\right]|n_1n_2m\rangle.
\end{equation}
For the remaining part in the right-hand-side of Eq. (\ref{exp}), we have
\begin{equation}
j_{1x}j_{2x}+j_{1y}j_{2y}=\frac{1}{2}\left(j_{1+}j_{2-}+j_{1-}j_{2+}\right).
\end{equation}
Since
\begin{equation}
\left(j_{1+}j_{2-}+j_{1-}j_{2+}\right)|n_1n_2m\rangle=\alpha_1|(n_1+1)(n_2-1)m\rangle+\alpha_2|(n_1-1)(n_2+1)m\rangle,
\end{equation}
with $\left(\alpha_1,\alpha_2\right)$ real coefficients, we get
\begin{equation}\label{int}
\sum_{\ell=|m|}^{n-1}\mathcal{B}^2(\ell)\left[n^2-1-\ell(\ell+1)\right]=\frac{n^2-1}{2}-\frac{1}{2}\left[m^2-\left(n_1-n_2\right)^2\right],
\end{equation}
where $\mathcal{B}(\ell)$ enables one to change from parabolic to spherical basis \cite{Park1960,Hughes1967,Arutyunyan1978,Mardoyan1985,Kibler1986}, \emph{i.e.}
\begin{equation}
|n_1n_2m\rangle=\sum_{\ell=|m|}^{n-1}\mathcal{B}(\ell)|n\ell m\rangle,
\end{equation}
with
\begin{eqnarray}\label{change}
\mathcal{B}(\ell)\equiv \mathcal{B}_{n,n_1,n_2,m}(\ell)&=&(-1)^{n_2+\frac{m-|m|}{2}+\ell}\sqrt{2\ell+1}\threej{\frac{n-1}{2}}{\frac{n-1}{2}}{\ell}{\frac{m-n_1+n_2}{2}}{\frac{m+n_1-n_2}{2}}{-m}\nonumber\\
&=&(-1)^{n_2+\frac{m-|m|}{2}+\ell}\sqrt{2\ell+1}\threej{\frac{n-1+m}{2}}{\frac{n-1-m}{2}}{\ell}{-\frac{q}{2}}{\frac{q}{2}}{0}\nonumber\\
&=&(-1)^{\ell-m}\frac{(n-m-1)!}{m!}\sqrt{\frac{(n_1+n_2)!(n_1+n_2+2m)!}{n_1!n_2!(n+\ell)!(\ell-m)!(n-\ell-1)!}}\nonumber\\
& &\times~_3F_2\left[\begin{array}{c}
\ell+m+1, -(\ell-m), -n_1\\
m+1, -(n-m-1)
\end{array};1\right].
\end{eqnarray}
The Clebsch-Gordan coefficient $\langle j_1m_1j_2m_2|j_3m_3\rangle$ is related to the $3jm$ symbol $\threej{j_1}{j_2}{j_3}{m_1}{m_2}{m_3}$ through the relation \cite{Varshalovich1988}:
\begin{equation}
\langle j_1m_1j_2m_2|j_3m_3\rangle=(-1)^{j_1-j_2+m_3}\threej{j_1}{j_2}{j_3}{m_1}{m_2}{m_3}.
\end{equation}
The two $3jm$ symbols on the first two lines of Eq. (\ref{change}) are equal because of the Regge symmetry \cite{Regge1958,Venkatesh1978}:
\begin{eqnarray}
\threej{j_1}{j_2}{j_3}{m_1}{m_2}{m_3}&=&\threej{j_1}{\frac{1}{2}\left(j_2+j_3+m_1\right)}{\frac{1}{2}\left(j_2+j_3-m_1\right)}{j_2-j_3}{\frac{1}{2}\left(j_3-j_2+m_1\right)+m_2}{\frac{1}{2}\left(j_3-j_2+m_1\right)+m_3}.\nonumber\\
& &
\end{eqnarray}
The two above expressions (last two equalities in Eq. (\ref{change}) in terms of $_3F_2$ hypergeometric functions \cite{Erdelyi1953,Slater1966}) were derived by Tarter \cite{Tarter1970}; they are not ``obvious'', in the sense that they do not coincide evidently with the usual expressions of the Clebsch-Gordan coefficients stemming from the summations by van der Warden, Fock, Racah, Wigner and Bandzaitis \cite{Warden1974,Fock1940,Racah1942,Wigner1959,Bandzaitis1964} and which correspond to Eqs. (21) to (27) pp. 240 and 241 of the reference book by Varshalovich, Moskalev and Khersonskii \cite{Varshalovich1988}. Its is worth mentioning that some special cases of the coefficients $\mathcal{B}(\ell)$ were pointed out by Hey, such as
\begin{equation}
\mathcal{B}_{n,n_1,n_2,\ell}(\ell)=\frac{1}{\ell!}\sqrt{\frac{(2\ell+1)!(n_1+\ell)!(n_2+\ell)!(n-\ell-1)!}{n_1!n_2!(n+\ell)!}},
\end{equation}
\begin{eqnarray}
\mathcal{B}(n-1)&=&\mathcal{B}_{n,n_1,n_2,m}(n-1)\nonumber\\
&=&(-1)^{n_2}(n-1)!\sqrt{\frac{(n_1+n_2)!(n_1+n_2+2m)!}{n_1!n_2!(2n-2)!(n_1+m)!(n_2+m)!}}\nonumber\\
& &
\end{eqnarray}
and
\begin{eqnarray}
\mathcal{B}(n-2)&=&\mathcal{B}_{n,n_1,n_2,m}(n-2)\nonumber\\
&=&(-1)^{n_2}(n_1-n_2)(n-1)!\nonumber\\
& &\times\sqrt{\frac{(2n-3)(n_1+n_2-1)!(n_1+n_2+2m-1)!}{n_1!n_2!(2n-2)!(n_1+m)!(n_2+m)!}}.
\end{eqnarray}
From Eq. (\ref{int}), we deduce
\begin{equation}\label{sr1}
\sum_{\ell=|m|}^{n-1}\mathcal{B}^2(\ell)\ell(\ell+1)=\frac{1}{2}\left[n^2-1+m^2-\left(n_1-n_2\right)^2\right],
\end{equation}
which was obtained by other means in Ref. \cite{Pain2021}. Calculating, in the same spirit, $\vl^{2p}$ with $p$ integer, in terms of operators $\vj_1$ and $\vj_2$, enables one to derive, at least in a recursive way or using a computer algebra system \cite{Mathematica}, expressions for
\begin{equation}
\sum_{\ell=|m|}^{n-1}\mathcal{B}^2(\ell)\left[\ell(\ell+1)\right]^p.
\end{equation}
For $m=0$, $n\gg 1$ and $n\gg\ell$, one has \cite{Drukarev1982,Rau1984a,Rau1984b,Fano1988}:
\begin{equation}
\mathcal{B}^2(\ell)\approx \frac{2\ell+1}{n}\exp\left[-\frac{\ell(\ell+1)}{n}\right].
\end{equation}

\section{The new sum rule}\label{sec5}

Using relation (\ref{change}) to change from parabolic to spherical basis, we get 
\begin{eqnarray}\label{chgba}
A_z^2|n_1n_2m\rangle&=&\sum_{\ell=|m|}^{n-1}(-1)^{\frac{1+n-n_1+n_2-m}{2}}\sqrt{2\ell+1}\nonumber\\
& &\times\threej{\frac{n-1}{2}}{\frac{n-1}{2}}{\ell}{\frac{m-n_1+n_2}{2}}{\frac{m+n_1-n_2}{2}}{-m}A_z^2|n\ell m\rangle.
\end{eqnarray}
According to Eq. (\ref{para}) we have
\begin{equation}\label{az2para}
A_z^2|n_1n_2m\rangle=(n_1-n_2)^2|n_1n_2m\rangle
\end{equation}
and using Eq. (\ref{nlm}):
\begin{eqnarray}
A_z^2|n\ell m\rangle&=&\left\{\frac{(\ell^2-m^2)(n^2-\ell^2)}{4\ell^2-1}\right.\nonumber\\
& &\left.+\frac{\left[(\ell+1)^2-m^2\right]\left[n^2-(\ell+1)^2\right]}{4(\ell+1)^2-1}\right\}|n\ell m\rangle\nonumber\\
& &+\left[\frac{(\ell^2-m^2)(n^2-\ell^2)}{(4\ell^2-1)}\right]^{1/2}\nonumber\\
& &\times\left\{\frac{\left[(\ell-1)^2-m^2\right]\left[n^2-(\ell-1)^2\right]}{4(\ell-1)^2-1}\right\}^{1/2}|n(\ell-2)m\rangle\nonumber\\
& &+\left\{\frac{\left[(\ell+2)^2-m^2\right]\left[n^2-(\ell+2)^2\right]}{4(\ell+2)^2-1}\right\}^{1/2}\nonumber\\
& &\times\left\{\frac{\left[(\ell+1)^2-m^2\right]\left[n^2-(\ell+1)^2\right]}{4(\ell+1)^2-1}\right\}^{1/2}|n(\ell+2)m\rangle.\nonumber\\
& &
\end{eqnarray}
Combining Eqs. (\ref{nlm}) and leads to (\ref{chgba})
\begin{eqnarray}
\langle n_1n_2 m|A_z^2|n_1n_2m\rangle&=&\sum_{\ell=|m|}^{n-1}\sum_{\ell'=|m|}^{n-1}\mathcal{B}(\ell)\mathcal{B}(\ell')\times\left\{\frac{(\ell^2-m^2)(n^2-\ell^2)}{4\ell^2-1}\right.\nonumber\\
& &\left.+\frac{\left[(\ell+1)^2-m^2\right]\left[n^2-(\ell+1)^2\right]}{4(\ell+1)^2-1}\right\}\langle n\ell'm|n\ell m\rangle\nonumber\\
& &+\sum_{\ell=|m|}^{n-1}\sum_{\ell'=|m|}^{n-1}\mathcal{B}(\ell)\mathcal{B}(\ell')\times\left[\frac{(\ell^2-m^2)(n^2-\ell^2)}{4\ell^2-1}\right]^{1/2}\nonumber\\
& &\times\left\{\frac{\left[(\ell-1)^2-m^2\right]\left[n^2-(\ell-1)^2\right]}{4(\ell-1)^2-1}\right\}^{1/2}\langle n\ell'm|n(\ell-2)m\rangle\nonumber\\
& &+\sum_{\ell=|m|}^{n-1}\sum_{\ell'=|m|}^{n-1}\mathcal{B}(\ell)\mathcal{B}(\ell')\times\left\{\frac{\left[(\ell+2)^2-m^2\right]\left[n^2-(\ell+2)^2\right]}{4(\ell+2)^2-1}\right\}^{1/2}\nonumber\\
& &\times\left\{\frac{\left[(\ell+1)^2-m^2\right]\left[n^2-(\ell+1)^2\right]}{4(\ell+1)^2-1}\right\}^{1/2}\langle n\ell'm|n(\ell+2)m\rangle\nonumber\\
& &
\end{eqnarray}
yielding
\begin{eqnarray}
\langle n_1n_2 m|A_z^2|n_1n_2m\rangle&=&\sum_{\ell=|m|}^{n-1}\mathcal{B}^2(\ell)\times\left\{\frac{(\ell^2-m^2)(n^2-\ell^2)}{4\ell^2-1}\right.\nonumber\\
& &\left.+\frac{\left[(\ell+1)^2-m^2\right]\left[n^2-(\ell+1)^2\right]}{4(\ell+1)^2-1}\right\}\nonumber\\
& &+\sum_{\ell=|m|}^{n-1}\mathcal{B}(\ell)\mathcal{B}(\ell-2)\left[\frac{(\ell^2-m^2)(n^2-\ell^2)}{4\ell^2-1}\right]^{1/2}\nonumber\\
& &\times\left\{\frac{\left[(\ell-1)^2-m^2\right]\left[n^2-(\ell-1)^2\right]}{4(\ell-1)^2-1}\right\}^{1/2}\nonumber\\
& &+\sum_{\ell=|m|}^{n-1}\mathcal{B}(\ell)\mathcal{B}(\ell+2)\left\{\frac{\left[(\ell+2)^2-m^2\right]\left[n^2-(\ell+2)^2\right]}{4(\ell+2)^2-1}\right\}^{1/2}\nonumber\\
& &\times\left\{\frac{\left[(\ell+1)^2-m^2\right]\left[n^2-(\ell+1)^2\right]}{4(\ell+1)^2-1}\right\}^{1/2}.
\end{eqnarray}
Using Eq. (\ref{az2para}), we finally obtain the sum rule
\begin{eqnarray}
\sum_{\ell=|m|}^{n-1}\mathcal{B}^2(\ell)\left\{\frac{(\ell^2-m^2)(n^2-\ell^2)}{4\ell^2-1}+\frac{\left[(\ell+1)^2-m^2\right]\left[n^2-(\ell+1)^2\right]}{4(\ell+1)^2-1}\right\}& &\nonumber\\
+\sum_{\ell=|m|}^{n-1}\mathcal{B}(\ell)\mathcal{B}(\ell-2)\left\{\frac{(\ell^2-m^2)(n^2-\ell^2)\left[(\ell-1)^2-m^2\right]\left[n^2-(\ell-1)^2\right]}{(4\ell^2-1)\left[4(\ell-1)^2-1\right]}\right\}^{1/2}& &\nonumber\\
+\sum_{\ell=|m|}^{n-1}\mathcal{B}(\ell)\mathcal{B}(\ell+2)\left\{\frac{\left[(\ell+2)^2-m^2\right]\left[n^2-(\ell+2)^2\right]\left[(\ell+1)^2-m^2\right]\left[n^2-(\ell+1)^2\right]}{(4\ell^2-1)\left[4(\ell+1)^2-1\right]}\right\}^{1/2}& &\nonumber\\
=(n_1-n_2)^2,& &
\end{eqnarray}
\emph{i.e.}
\begin{eqnarray}\label{sr2}
& &\sum_{\ell=|m|}^{n-1}(2\ell+1)\left\{\frac{(\ell^2-m^2)(n^2-\ell^2)}{4\ell^2-1}+\frac{\left[(\ell+1)^2-m^2\right]\left[n^2-(\ell+1)^2\right]}{4(\ell+1)^2-1}\right\}\nonumber\\
& &\times\threej{\frac{n-1}{2}}{\frac{n-1}{2}}{\ell}{\frac{m-n_1+n_2}{2}}{\frac{m+n_1-n_2}{2}}{-m}^2\nonumber\\
& &+\sum_{\ell=|m|}^{n-1}\sqrt{(2\ell+1)(2\ell-3)}\left\{\frac{(\ell^2-m^2)(n^2-\ell^2)\left[(\ell-1)^2-m^2\right]\left[n^2-(\ell-1)^2\right]}{(4\ell^2-1)\left[4(\ell-1)^2-1\right]}\right\}^{1/2}\nonumber\\
& &\times\threej{\frac{n-1}{2}}{\frac{n-1}{2}}{\ell}{\frac{m-n_1+n_2}{2}}{\frac{m+n_1-n_2}{2}}{-m}\threej{\frac{n-1}{2}}{\frac{n-1}{2}}{\ell-2}{\frac{m-n_1+n_2}{2}}{\frac{m+n_1-n_2}{2}}{-m}\nonumber\\
& &+\sum_{\ell=|m|}^{n-1}\sqrt{(2\ell+1)(2\ell+5)}\nonumber\\
& &\times\left\{\frac{\left[(\ell+2)^2-m^2\right]\left[n^2-(\ell+2)^2\right]\left[(\ell+1)^2-m^2\right]\left[n^2-(\ell+1)^2\right]}{(4\ell^2-1)\left[4(\ell+1)^2-1\right]}\right\}^{1/2}\nonumber\\
& &\times\threej{\frac{n-1}{2}}{\frac{n-1}{2}}{\ell}{\frac{m-n_1+n_2}{2}}{\frac{m+n_1-n_2}{2}}{-m}\threej{\frac{n-1}{2}}{\frac{n-1}{2}}{\ell+2}{\frac{m-n_1+n_2}{2}}{\frac{m+n_1-n_2}{2}}{-m}\nonumber\\
& &=(n_1-n_2)^2.
\end{eqnarray}

\section{Connection with the diamagnetism of the hydrogen atom in the low-field limit}\label{sec6}

The Hamiltonian of the hydrogen atom subject to a weak magnetic field reads
\begin{equation}\label{hdia}
H=\frac{p^2}{2}-\frac{1}{r}+\frac{\gamma}{2}L_z+\frac{\gamma^2}{8}\left(x^2+y^2\right),
\end{equation}
where $\gamma$ is the ratio of the cyclotron frequency to the Rydberg constant. The treatment of the gyromagnetic term $\gamma L_z/2$ is straightforward as $L_z$ is a constant of motion (see Eq. (\ref{lz})). The treatment of the diamagnetic term $\gamma^2\left(x^2+y^2\right)/8$ is more tedious as the total Hamiltonian of Eq. (\ref{hdia}) does not separate. The first attempt to derive the adiabatic invariant in the low-field regime was made by Labarthe \cite{Labarthe1981} who resorted to the Fermi replacement $\vr\rightarrow 3n\va/2$. Two important contributions gave a complete description of the phenomenon in the low-field limit, through the discovery of an adiabatic invariant related to the diamagnetic interaction. The first derivation was first given by Solov'ev \cite{Solovev1982} who resorted to classical perturbation theory. He showed that $\rho^2=x^2+y^2$ can be expressed in a given shell as a linear function of $4A^2-5A_z$. Herrick \cite{Herrick1982} obtained a similar expression using the momentum representation of the Coulomb problem on the Fock hypersphere of $SO(4)$. Although these derivations solve the problem in the low-field limit, they require quite long and cumbersome calculations. Moreover, they hinder the basic physical meaning of the results \cite{Gay1983}. Delande and Gay \cite{Delande1984} presented a completely different analysis based on symmetry considerations. Applying the above generalized vectorial model to the diamagnetic interaction, which has close connections with the non-invariance algebra of the Coulomb problem $SO(4,1)$ \cite{Delande1986,Grozdanov1986} mentioned in Sec. \ref{sec1}, the authors derived the effective diamagnetic Hamiltonian up to second order in the magnetic field $B$ in terms of the generators $(\vj_1,\vj_2)$ of the $SO(4)$ Lie algebra for the Coulomb problem. The result is
\begin{equation}
H=H_0+H_1+H_2
\end{equation}
with
\begin{eqnarray}
H_1&=&\frac{\gamma^2n^2}{16}\left(3n^2+1-4j_{1z}^2-4j_{2z}^2+4j_{1z}j_{2z}-4j_{1+}j_{2-}-4j_{1-}j_{2+}\right)\nonumber\\
&=&\frac{\gamma^2n^2}{16}\left(n^2+3+L_z^2+4A^2-5A_z^2\right)
\end{eqnarray}
and
\begin{eqnarray}
H_2&=&\left(\frac{\gamma^2}{8}\right)^2\frac{n^6}{48}\left\{-223n^4-598n^2-27+192\left(j_{1z}^4+j_{2z}^4\right)\right.\nonumber\\
& &\left.+144j_{1z}^2j_{2z}^2-(176n^2+752)j_{1z}j_{2z}\right.\nonumber\\
& &\left.+\left(j_{1z}^2+j_{2z}^2\right)\left(-32j_{1z}j_{2z}+284n^2+372\right)+8\left(j_{1+}j_{2-}+j_{1-}j_{2+}\right)\right.\nonumber\\
& &\times\left[53n^2+153+20\left(j_{1z}^2+j_{2z}^2\right)-12j_{1z}j_{2z}\right]\nonumber\\
& &\left.+208\left(j_{1+}j_{2-}-j_{1-}j_{2+}\right)\left(j_{1z}-j_{2z}\right)+48\left(j_{1+}^2j_{2-}^2+j_{1-}^2j_{2+}^2\right)\right\}.\nonumber\\
& &
\end{eqnarray}
Higher orders can be deduced exactly with the use of computer programs for algebraic manipulations \cite{Mathematica}, as for the sum rule of Sec. \ref{sec4} generalized to powers of $\ell(\ell+1)$.
\section{Stark effect, higher powers of $A_z$ and further sum rules}\label{sec7}

The Hamiltonian of a hydrogen atom in an electric field $\vF$ is
\begin{equation}
H=\frac{p^2}{2}-\frac{1}{r}-\vF.\vr.
\end{equation}
The intensity of the field can depend on time (this corresponds to the so-called AC - Alternative Current - Stark effect). On a basis set made of all zero-field wavefunctions having the same principal quantum number $n$, the Hamiltonian can be replaced by the projected Hamiltonian \cite{Englefield1972}:
\begin{equation}
H_n=-\frac{1}{2n^2}-\frac{3}{2}\vF.\va,
\end{equation}
with
\begin{equation}
\va=n\left[\vp.(\vr.\vp)-r\left(p^2-\frac{1}{r}\right)\right].
\end{equation}
Taking into account only the field-dependent part of $H$, the time-evolution operator verifies
\begin{equation}
i\frac{\partial \mathcal{U}_n}{\partial t}=-\frac{3}{2}n\left(\vF.\va\right)\mathcal{U}_n,
\end{equation}
and $U_n$ can be viewed as an approximation for the modeling of the angular momentum shifts induced by the electric field.

The evolution operator describing the overall transition from initial to final states can be described by a single $SO(4)$ rotation. Using Biedenharn's \cite{Biedenharn1961} parametrization of rotation in four dimensions we have then
\begin{equation}
\mathcal{U}_n=e^{i\alpha_1L_z}e^{i\alpha_2L_y}e^{i\alpha_3L_z}e^{i\alpha_4A_z}e^{i\alpha_5L_y}e^{i\alpha_6L_z}
\end{equation}
where $\alpha_1$, $\alpha_2$, $\alpha_3$, $\alpha_5$ and $\alpha_6$ are Euler angles for ordinary three-dimensional rotations, which do not change $\ell$. The net angular momentum transfer in the hydrogenic model is therefore described by the single time-dependent parameter $\alpha_4$. The average transition probability from an initial state $n\ell$ to all final states of $n\ell'$ is

\begin{equation}
\mathcal{P}(\ell,\ell')=\frac{1}{2\ell+1}\sum_{m,m'}\left|\langle n\ell' m'|\mathcal{U}_n|n\ell m\rangle\right|^2,
\end{equation}
where the summation over $m$ and $m'$ extends over the $(2\ell+1)$-fold and $(2\ell'+1)$-fold three-dimensional angular-momentum substates, respectively. $\mathcal{P}(\ell,\ell')$ is independent of the three-dimensional rotations and reduces to
\begin{equation}\label{expa}
\mathcal{P}(\ell,\ell')=\frac{1}{2\ell+1}\sum_{m}|\langle n\ell'm|e^{i\chi A_z}|n\ell m\rangle|^2,
\end{equation}
where $\chi$, characteristic of the cumulative field seen by the atom, has been substituted for $\alpha_4$. One has \cite{Herrick1977}:
\begin{equation}
\mathcal{P}(\ell,\ell')=(2\ell'+1)\sum_{m,q,q'}\mathcal{C}(q\ell m)\mathcal{C}(q'\ell m)\mathcal{C}(q\ell' m)\mathcal{C}(q'\ell' m)\cos\left[\chi(q-q')\right]
\end{equation} 
with
\begin{equation}
\mathcal{C}(q\ell m)=\threej{\frac{n-1}{2}}{\frac{n-1}{2}}{\ell}{\frac{m-q}{2}}{\frac{m+q}{2}}{-m},
\end{equation}
where $q=n_1-n_2$. For most values of the field accessible in experiments, $\mathcal{P}(\ell,\ell')$ is a rapidly oscillating function of time ($\chi=3nt/2$) and the time-averaged angular-momentum transfer can be approximated by the diagonal part only ($q=q'$):
The latter equation can be transformed into
\begin{equation}
\mathcal{P}(\ell,\ell')\approx\bar{\mathcal{P}}(\ell,\ell')=(2\ell'+1)\sum_{m,q}\mathcal{C}^2(q\ell m)\mathcal{C}^2(q\ell' m),
\end{equation}
which can be expressed as
\begin{equation}
\bar{\mathcal{P}}(\ell,\ell')=(2\ell'+1)\sum_{j=0}^{\ell-\ell'}\sixj{\ell}{\ell'}{j}{\frac{n-1}{2}}{\frac{n-1}{2}}{\frac{n-1}{2}}^2.
\end{equation}
Such a quantity is difficult to evaluate (it is often referred to as an ``unusual sum rule''), since it does not involve the weighting factor $(2j+1)$. However, specifying the value of $\ell$, one can obtain new sum rules; indeed one has for instance
\begin{equation}
\bar{\mathcal{P}}(0,\ell)=\frac{1}{n}
\end{equation}
and
\begin{equation}
\bar{\mathcal{P}}(1,\ell)=\frac{n^2\left[4\ell(\ell+1)-1\right]-2(\ell+1)+1}{n\left(n^2-1\right)(2\ell-1)(2\ell+3)}.
\end{equation}
An expression for $\mathcal{P}(2,\ell)$ was published by Herrick \cite{Herrick1977}. In equation (\ref{expa}), the exponential function can be expanded as a power series in $A_z$; therefore, it would be interesting to be able to calculate
\begin{equation}
\langle n\ell m|A_z^k|n\ell m\rangle,
\end{equation}
for any integer value of $k$. As we have seen in section \ref{sec3} for $A_z^2$, the ket $|n\ell m\rangle$ is transformed into a linear combination of $|n(\ell-2) m\rangle$, $|n\ell m\rangle$ and $|n(\ell+2) m\rangle$. In the same way, one finds that $A_z^3$ transforms $|n\ell m\rangle$ into a linear combination of $|n(\ell-3) m\rangle$, $|n(\ell-1) m\rangle$, $|n(\ell+1) m\rangle$ and $|n(\ell+3) m\rangle$. By induction, it is easy to prove that $A_z^k$ transforms $|n\ell m\rangle$ into a linear combination of $|n(\ell-k) m\rangle, |n(\ell-k+2) m\rangle, |n(\ell-k+4) m\rangle, \cdots |n(\ell+k-4) m\rangle, |n(\ell+k-2) m\rangle$ and $|n(\ell+k) m\rangle$. In parabolic coordinate, the result is very simple:
\begin{equation}
\langle n_1n_2m|A_z^k|n_1n_2 m\rangle=(n_1-n_2)^k.
\end{equation}
Let us define 
\begin{equation}
\beta_{n\ell}=\left[\frac{\left(n^2-\ell^2\right)\left(\ell^2-m^2\right)}{4\ell^2-1}\right]^{1/2},
\end{equation}
which enables us to write
\begin{equation}
A_z|n\ell m\rangle=\beta_{n(\ell+1)}|n(\ell+1)m\rangle+\beta_{n\ell}|n(\ell-1)m\rangle.
\end{equation}
The sum rule (\ref{sr2}) reads
\begin{eqnarray}
& &\sum_{\ell=|m|}^{n-1}\mathcal{B}(\ell)\mathcal{B}(\ell-2)\beta_{n\ell}\beta_{n(\ell-1)}+\sum_{\ell=|m|}^{n-1}\mathcal{B}^2(\ell)\left[\beta_{n\ell}^2+\beta_{n(\ell+1)}^2\right]\nonumber\\
& &+\sum_{\ell=|m|}^{n-1}\mathcal{B}(\ell)\mathcal{B}(\ell+2)\beta_{n(\ell+1)}\beta_{n(\ell+2)}=(n_1-n_2)^2
\end{eqnarray}
or
\begin{eqnarray}
& &\sum_{\ell=|m|}^{n-1}\sqrt{(2\ell+1)(2\ell-3)}~\beta_{n(\ell-1)}\beta_{n\ell}\nonumber\\
& &\times\threej{\frac{n-1}{2}}{\frac{n-1}{2}}{\ell}{\frac{m-n_1+n_2}{2}}{\frac{m+n_1-n_2}{2}}{-m}\threej{\frac{n-1}{2}}{\frac{n-1}{2}}{\ell-2}{\frac{m-n_1+n_2}{2}}{\frac{m+n_1-n_2}{2}}{-m}\nonumber\\
& &+\sum_{\ell=|m|}^{n-1}(2\ell+1)\left[\beta_{n\ell}^2+\beta_{n(\ell+1)}^2\right]\threej{\frac{n-1}{2}}{\frac{n-1}{2}}{\ell}{\frac{m-n_1+n_2}{2}}{\frac{m+n_1-n_2}{2}}{-m}^2\nonumber\\
& &+\sum_{\ell=|m|}^{n-1}\sqrt{(2\ell+1)(2\ell+5)}~\beta_{n(\ell+2)}\nonumber\\
& &\times\threej{\frac{n-1}{2}}{\frac{n-1}{2}}{\ell}{\frac{m-n_1+n_2}{2}}{\frac{m+n_1-n_2}{2}}{-m}\threej{\frac{n-1}{2}}{\frac{n-1}{2}}{\ell+2}{\frac{m-n_1+n_2}{2}}{\frac{m+n_1-n_2}{2}}{-m}\beta_{n(\ell+1)}\nonumber\\
& &=(n_1-n_2)^2.
\end{eqnarray}
We have also
\begin{eqnarray}
A_z^3|n\ell m\rangle&=&\beta_{n(\ell-2)}\beta_{n(\ell-1)}\beta_{n\ell}|n(\ell-3)m\rangle\nonumber\\
& &+\beta_{n\ell}\left[\beta_{n(\ell-1)}^2+\beta_{n\ell}^2+\beta_{n(\ell+1)}^2\right]|n(\ell-1)m\rangle\nonumber\\
& &+\beta_{n(\ell+1)}\left[\beta_{n\ell}^2+\beta_{n(\ell+1)}^2+\beta_{n(\ell+2)}^2\right]|n(\ell+1)m\rangle\nonumber\\
& &+\beta_{n(\ell+1)}\beta_{n(\ell+2)}\beta_{n(\ell+3)}|n(\ell+3)m\rangle
\end{eqnarray}
yielding the sum rule
\begin{eqnarray}
& &\sum_{\ell=|m|}^{n-1}\mathcal{B}(\ell)\mathcal{B}(\ell-3)\beta_{n(\ell-2)}\beta_{n(\ell-1)}\beta_{n\ell}\nonumber\\
& &+\sum_{\ell=|m|}^{n-1}\mathcal{B}(\ell)\mathcal{B}(\ell-1)\beta_{n\ell}\left[\beta_{n(\ell-1)}^2+\beta_{n\ell}^2+\beta_{n(\ell+1)}^2\right]\nonumber\\
& &+\sum_{\ell=|m|}^{n-1}\mathcal{B}(\ell)\mathcal{B}(\ell+1)\beta_{n(\ell+1)}\left[\beta_{n\ell}^2+\beta_{n(\ell+1)}^2+\beta_{n(\ell+2)}^2\right]\nonumber\\
& &+\sum_{\ell=|m|}^{n-1}\mathcal{B}(\ell)\mathcal{B}(\ell+3)\beta_{n(\ell+1)}\beta_{n(\ell+2)}\beta_{n(\ell+3)}\nonumber\\
&&=\left(n_1-n_2\right)^3,
\end{eqnarray}
\emph{i.e.}
\begin{eqnarray}\label{sr3}
& &\sum_{\ell=|m|}^{n-1}\sqrt{(2\ell+1)(2\ell-5)}~\beta_{n(\ell-2)}\beta_{n(\ell-1)}\beta_{n\ell}\nonumber\\
& &\times\threej{\frac{n-1}{2}}{\frac{n-1}{2}}{\ell}{\frac{m-n_1+n_2}{2}}{\frac{m+n_1-n_2}{2}}{-m}\threej{\frac{n-1}{2}}{\frac{n-1}{2}}{\ell-3}{\frac{m-n_1+n_2}{2}}{\frac{m+n_1-n_2}{2}}{-m}\nonumber\\
& &+\sum_{\ell=|m|}^{n-1}\sqrt{(4\ell^2-1})~\beta_{n\ell}\left[\beta_{n(\ell-1)}^2+\beta_{n\ell}^2+\beta_{n(\ell+1)}^2\right]\nonumber\\
& &\times\threej{\frac{n-1}{2}}{\frac{n-1}{2}}{\ell}{\frac{m-n_1+n_2}{2}}{\frac{m+n_1-n_2}{2}}{-m}\threej{\frac{n-1}{2}}{\frac{n-1}{2}}{\ell-1}{\frac{m-n_1+n_2}{2}}{\frac{m+n_1-n_2}{2}}{-m}\nonumber\\
& &+\sum_{\ell=|m|}^{n-1}\sqrt{(2\ell+1)(2\ell+3)}~\beta_{n(\ell+1)}\left[\beta_{n\ell}^2+\beta_{n(\ell+1)}^2+\beta_{n(\ell+2)}^2\right]\nonumber\\
& &\times\threej{\frac{n-1}{2}}{\frac{n-1}{2}}{\ell}{\frac{m-n_1+n_2}{2}}{\frac{m+n_1-n_2}{2}}{-m}\threej{\frac{n-1}{2}}{\frac{n-1}{2}}{\ell+1}{\frac{m-n_1+n_2}{2}}{\frac{m+n_1-n_2}{2}}{-m}\nonumber\\
& &+\sum_{\ell=|m|}^{n-1}\sqrt{(2\ell+1)(2\ell+7)}~\beta_{n(\ell+1)}\beta_{n(\ell+2)}\beta_{n(\ell+3)}\nonumber\\
& &\times\threej{\frac{n-1}{2}}{\frac{n-1}{2}}{\ell}{\frac{m-n_1+n_2}{2}}{\frac{m+n_1-n_2}{2}}{-m}\threej{\frac{n-1}{2}}{\frac{n-1}{2}}{\ell+3}{\frac{m-n_1+n_2}{2}}{\frac{m+n_1-n_2}{2}}{-m}\nonumber\\
& &=\left(n_2-n_1\right)^3.
\end{eqnarray}
In the same spirit, we have also
\begin{equation}
\begin{tabular}{ll}
$A_z^4|n\ell m\rangle=$ & $\beta_{n(\ell-3)}\beta_{n(\ell-2)}\beta_{n(\ell-1)}\beta_{n\ell}|n(\ell-4)m\rangle$\\
& $+\beta_{n(\ell-1)}\beta_{n\ell}\left[\beta_{n(\ell-2)}^2+\beta_{n(\ell-1)}^2+\beta_{n\ell}^2+\beta_{n(\ell+1)}^2\right]|n(\ell-2)m\rangle$\\
& $+\left\{\beta_{n(\ell+1)}^2\left[\beta_{n\ell}^2+\beta_{n(\ell+1)}^2+\beta_{n(\ell+2)}^2\right]\right.$\\
& $+\left.\beta_{n\ell}^2\left[\beta_{n(\ell-1)}^2+\beta_{n\ell}^2+\beta_{n(\ell+1)}^2\right]\right\}|n\ell m\rangle$\\
& $+\beta_{n(\ell+1)}\beta_{n(\ell+2)}\left[\beta_{n\ell}^2+\beta_{n(\ell+1)}^2+\beta_{n(\ell+2)}^2+\beta_{n(\ell+3)}^2\right]|n(\ell+2)m\rangle$\\
& $+\beta_{n(\ell+1)}\beta_{n(\ell+2)}\beta_{n(\ell+3)}\beta_{n(\ell+4)}|n(\ell+4)m\rangle$
\end{tabular}
\end{equation}
yielding the sum rule
\begin{eqnarray}
& &\sum_{\ell=|m|}^{n-1}\mathcal{B}(\ell)\mathcal{B}(\ell-4)\beta_{n(\ell-3)}\beta_{n(\ell-2)}\beta_{n(\ell-1)}\beta_{n\ell}\nonumber\\
& &+\sum_{\ell=|m|}^{n-1}\mathcal{B}(\ell)\mathcal{B}(\ell-2)\beta_{n(\ell-1)}\beta_{n\ell}\left[\beta_{n(\ell-2)}^2+\beta_{n(\ell-1)}^2+\beta_{n\ell}^2+\beta_{n(\ell+1)}^2\right]\nonumber\\
& &+\sum_{\ell=|m|}^{n-1}\mathcal{B}^2(\ell)\left\{\beta_{n(\ell+1)}^2\left[\beta_{n\ell}^2+\beta_{n(\ell+1)}^2+\beta_{n(\ell+2)}^2\right]+\beta_{n\ell}^2\left[\beta_{n(\ell-1)}^2+\beta_{n\ell}^2+\beta_{n(\ell+1)}^2\right]\right\}\nonumber\\
& &+\sum_{\ell=|m|}^{n-1}\mathcal{B}(\ell)\mathcal{B}(\ell+2)\beta_{n(\ell+1)}\beta_{n(\ell+2)}\left[\beta_{n\ell}^2+\beta_{n(\ell+1)}^2+\beta_{n(\ell+2)}^2+\beta_{n(\ell+3)}^2\right]\nonumber\\
& &+\sum_{\ell=|m|}^{n-1}\mathcal{B}(\ell)\mathcal{B}(\ell+4)\beta_{n(\ell+1)}\beta_{n(\ell+2)}\beta_{n(\ell+3)}\beta_{n(\ell+4)}\nonumber\\
& &=\left(n_2-n_1\right)^4,
\end{eqnarray}
or alternatively
\begin{eqnarray}\label{sr4}
& &\sum_{\ell=|m|}^{n-1}\sqrt{(2\ell+1)(2\ell-7)}\beta_{n(\ell-3)}~\beta_{n(\ell-2)}\beta_{n(\ell-1)}\beta_{n\ell}\nonumber\\
& &\times\threej{\frac{n-1}{2}}{\frac{n-1}{2}}{\ell}{\frac{m-n_1+n_2}{2}}{\frac{m+n_1-n_2}{2}}{-m}\threej{\frac{n-1}{2}}{\frac{n-1}{2}}{\ell-4}{\frac{m-n_1+n_2}{2}}{\frac{m+n_1-n_2}{2}}{-m}\nonumber\\
& &+\sum_{\ell=|m|}^{n-1}\sqrt{(2\ell+1)(2\ell-3)}~\beta_{n(\ell-1)}\beta_{n\ell}\left[\beta_{n(\ell-2)}^2+\beta_{n(\ell-1)}^2+\beta_{n\ell}^2+\beta_{n(\ell+1)}^2\right]\nonumber\\
& &\times\threej{\frac{n-1}{2}}{\frac{n-1}{2}}{\ell}{\frac{m-n_1+n_2}{2}}{\frac{m+n_1-n_2}{2}}{-m}\threej{\frac{n-1}{2}}{\frac{n-1}{2}}{\ell-2}{\frac{m-n_1+n_2}{2}}{\frac{m+n_1-n_2}{2}}{-m}\nonumber\\
& &+\sum_{\ell=|m|}^{n-1}(2\ell+1)~\left\{\beta_{n(\ell+1)}^2\left[\beta_{n\ell}^2+\beta_{n(\ell+1)}^2+\beta_{n(\ell+2)}^2\right]+\beta_{n\ell}^2\left[\beta_{n(\ell-1)}^2+\beta_{n\ell}^2+\beta_{n(\ell+1)}^2\right]\right\}\nonumber\\
& &\times\threej{\frac{n-1}{2}}{\frac{n-1}{2}}{\ell}{\frac{m-n_1+n_2}{2}}{\frac{m+n_1-n_2}{2}}{-m}^2\nonumber\\
& &+\sum_{\ell=|m|}^{n-1}\sqrt{(2\ell+1)(2\ell+5)}~\beta_{n(\ell+1)}\beta_{n(\ell+2)}\left[\beta_{n\ell}^2+\beta_{n(\ell+1)}^2+\beta_{n(\ell+2)}^2+\beta_{n(\ell+3)}^2\right]\nonumber\\
& &\times\threej{\frac{n-1}{2}}{\frac{n-1}{2}}{\ell}{\frac{m-n_1+n_2}{2}}{\frac{m+n_1-n_2}{2}}{-m}\threej{\frac{n-1}{2}}{\frac{n-1}{2}}{\ell+2}{\frac{m-n_1+n_2}{2}}{\frac{m+n_1-n_2}{2}}{-m}\nonumber\\
& &+\sum_{\ell=|m|}^{n-1}\sqrt{(2\ell+1)(2\ell+9)}~\beta_{n(\ell+1)}\beta_{n(\ell+2)}\beta_{n(\ell+3)}\beta_{n(\ell+4)}\nonumber\\
& &\times\threej{\frac{n-1}{2}}{\frac{n-1}{2}}{\ell}{\frac{m-n_1+n_2}{2}}{\frac{m+n_1-n_2}{2}}{-m}\threej{\frac{n-1}{2}}{\frac{n-1}{2}}{\ell+4}{\frac{m-n_1+n_2}{2}}{\frac{m+n_1-n_2}{2}}{-m}\nonumber\\
& &=\left(n_2-n_1\right)^4.
\end{eqnarray}

\section{Example and perspectives}

For instance, in the case $n=9$ and $m=4$, and the parabolic quantum numbers $n_1=3$ and $n_2=1$ (yielding $q=n_1-n_2=2$), we have indicated in Table \ref{tab1} the values of the sums given in the left-hand sides of Eqs. (\ref{sr1}), (\ref{sr2}), (\ref{sr3}) and (\ref{sr4}) respectively.
\begin{table}[]
    \centering
    \begin{tabular}{|c|c|c|}\hline
 Sum & Analytical value & Numerical value  \\\hline
 $\mathscr{S}_1$=Left-hand side of Eq. (\ref{sr1}) & $\frac{1}{2}\left[n^2-1+m^2-(n_1-n_2)^2\right]$ & 46\\\hline
 $\mathscr{S}_2$=Left-hand side of Eq. (\ref{sr2}) & $(n_1-n_2)^2$ & 4 \\\hline
 $\mathscr{S}_3$=Left-hand side of Eq. (\ref{sr3}) & $(n_1-n_2)^3$ & 8 \\\hline
 $\mathscr{S}_4$=Left-hand side of Eq. (\ref{sr4}) & $(n_1-n_2)^4$ & 16 \\\hline
 \end{tabular}
    \caption{Illustration of the sum rules (\ref{sr1}), (\ref{sr2}), (\ref{sr3}) and (\ref{sr4}) in the case $n=9$, $m=4$, $n_1=3$ and $n_2=1$ (yielding $q=n_1-n_2=2$).}
    \label{tab1}
\end{table}
$\mathscr{S}_1$ is given by the left-hand side of Eq. (\ref{sr1}) and reads
\begin{equation}
\mathscr{S}_1=\sum_{\ell=4}^{8}(2\ell+1)\threej{6}{2}{\ell}{-1}{1}{0}^2\ell(\ell+1),
\end{equation}
while $\mathscr{S}_2$ is provided by the left-hand side of Eq. (\ref{sr2}):
\begin{eqnarray}
&&\mathscr{S}_2=\sum_{\ell=4}^{8}(2\ell+1)\left\{\frac{(\ell^2-16)(81-\ell^2)}{4\ell^2-1}+\frac{\left[(\ell+1)^2-16\right]\left[81-(\ell+1)^2\right]}{4(\ell+1)^2-1}\right\}\nonumber\\
& &\;\;\;\;\;\;\;\;\times\threej{6}{2}{\ell}{-1}{1}{0}^2\nonumber\\
& &\;\;\;\;\;\;\;\;+\sum_{\ell=4}^{8}\sqrt{(2\ell+1)(2\ell-3)}\left\{\frac{(\ell^2-16)(81-\ell^2)\left[(\ell-1)^2-16\right]\left[81-(\ell-1)^2\right]}{(4\ell^2-1)\left[4(\ell-1)^2-1\right]}\right\}^{1/2}\nonumber\\
& &\;\;\;\;\;\;\;\;\times\threej{6}{2}{\ell}{-1}{1}{0}\threej{6}{2}{\ell-2}{-1}{1}{0}\nonumber\\
& &\;\;\;\;\;\;\;\;+\sum_{\ell=4}^{8}\sqrt{(2\ell+1)(2\ell+5)}\nonumber\\
& &\;\;\;\;\;\;\;\;\times\left\{\frac{\left[(\ell+2)^2-16\right]\left[81-(\ell+2)^2\right]\left[(\ell+1)^2-16\right]\left[81-(\ell+1)^2\right]}{(4\ell^2-1)\left[4(\ell+1)^2-1\right]}\right\}^{1/2}\nonumber\\
& &\;\;\;\;\;\;\;\;\times\threej{6}{2}{\ell}{-1}{1}{0}\threej{6}{2}{\ell+2}{-1}{1}{0}.
\end{eqnarray}
In the same way, for $\mathscr{S}_3$ and $\mathscr{S}_4$, one just has to replace $n$, $n_1$, $n_2$ and $m$ by their respective values in the left-hand sides of Eqs. (\ref{sr3}) and (\ref{sr4}).

The method described here is based on the dynamical symmetry of the hydrogen atom and the preservation of the Runge-Lenz-Pauli vector. The basic ideas, however, could stimulate investigations related to other groups. They may for instance find applications within the quantum group $SU_q(2)$ \cite{Macfarlane1989,Nomura1990}. The general scheme underlying this work is the calculation of matrix elements of an invariant operator. Therefore, whatever the groups involved, the most difficult point is to find such an invariant (Casimir operator, tensor, etc.) and calculate its matrix element in different basis. 

A quasi-spin tensor decomposition of the two-nucleon interaction determines the seniority conserving rotationally invariant two-body interaction in a single-$j$ shell, playing a major role in nuclear physics. Such interactions define solvable shell model Hamiltonians for which the unitary symplectic algebra $USp(2j+1)$ gives a complete ensemble of quantum numbers for a subset of states. The latter are uniquely defined by seniority and angular momentum which means that the model has a partial dynamical symmetry \cite{Rosensteel2003}. 

Barut and Wilson proposed a systematic derivation of various relations and identities among the Clebsch-Gordan coefficients and the representation functions of $SO(4)$ and $SO(2,1)$. These relations are essential in works involving the matrix elements of arbitrary group elements in higher noncompact groups such as $O(4,2)$. The latter contains most of the physical groups like the Lorentz group $O(3,1)$ (important for the description of the continuum states of the hydrogen atom), $O(3,2)$, $O(2,1)$, $E(3)$, etc. \cite{Barut1976}. In the case of a charge placed in the fields of a magnetic monopole, there is a privileged space direction and the corresponding dynamical symmetry is $SO(4,2)$. 

As concerns the Stark effect already mentioned \cite{Herrick1975}, when the hydrogen atom is subject to an electric field, the dynamical symmetry is broken and the Runge-Lenz vector is no more an invariant. Nevertheless, Redmond could show that the following quantity
\begin{equation}
\mathscr{R}=\va.\vF-\frac{1}{2}(\vr \times \vF)^2 
\end{equation}
is a dynamical invariant \cite{Redmond1964}. We plan to apply the techniques described in the present work to the Redmond invariant.

\section{Conclusion}

We presented sum rules (Eqs. (\ref{sr2}), (\ref{sr3}) and (\ref{sr4})) for Clebsch-Gordan coefficients in the framework of $SO(4)$ group-theoretical description of the hydrogen atom. An alternative derivation of a sum rule previously obtained was first proposed in order to introduce the formalism and methodology. The latter identity is the first $(k=1)$  of a family of sum rules involving $[\ell(\ell+1)]^k$ in addition to the $(2\ell+1)$ factor. Such sum rules are useful in order to make the connection between average values of powers of the radial vector in spherical and parabolic coordinates, and to determine the rotational energy-weighted moments in rotational molecular spectra. More generally, such identities can help checking analytical or numerical calculations involving a change of variables from spherical to parabolic coordinates, such as the Stark effect or the semianalytic study of diamagnetism in a degenerate hydrogenic manifold.

The new identities are obtained using properties of the Runge-Lenz-Pauli vector, and in particular expressing the matrix element of the powers 2, 3 and 4 of its last component both in spherical and parabolic basis. The sum rules are unusual, in the sense that the weighting coefficients in front of the squared Clebsch-Gordan coefficients are ratios of polynomials in $\ell$. The sums involve products of two binomial coefficients differing only by one parameter and are equal to $(n_1-n_2)^p$. For instance for $p=2$, the above mentioned pairs of differing parameters are $(\ell,\ell)$, $(\ell,\ell+2)$ and $(\ell,\ell-2)$. For $p=3$, they are $(\ell,\ell+1)$, $(\ell,\ell-1)$, $(\ell,\ell+3)$ and $(\ell,\ell-3)$. For $p=4$, they are $(\ell,\ell)$, $(\ell,\ell+2)$, $(\ell,\ell-2)$, $(\ell,\ell+4)$, $(\ell,\ell-4)$. More generally, if $p$ is even, ($p=2r$), the pairs of differing parameters are ($\ell,\ell\pm 2j$), $j$ ranging from 0 to $r$ and if $p$ is odd ($p=2r+1$), from ($\ell,\ell\pm 2j\pm1$), $j$ ranging from 0 to $r$.

Besides their strong connections with Stark and Zeeman effects, the elementary calculations presented here may certainly be extended and yield further various developments of sum identities involving $3jm$ or Clebsch-Gordan coefficients. 


\end{document}